\documentclass[preprint]{aastex}
\newcommand\pcc{\;{\rm cm}^{-3}}

\newcommand\kms{\; {\rm km}\;{\rm s}^{-1}}
\newcommand\ergs{\; {\rm erg}\;{\rm s}^{-1}}

\newcommand\mh{\; m_{\rm H}}
\newcommand\cm{\;{\rm cm}}

\newcommand\Myr{\;{\rm Myr}}

\newcommand\pc{\;{\rm pc}}

\newcommand\Punit{\pcc\,{\rm K}}

\newcommand\Kel{\;{\rm K}}

\newcommand\simgt{\lower.5ex\hbox{$\; \buildrel > \over \sim \;$}}
\newcommand\simlt{\lower.5ex\hbox{$\; \buildrel < \over \sim \;$}}
\newcommand\pderiv[2]{\frac{\partial {#1}}{\partial {#2}}}

\newcommand\rbrackets[1]{\left({#1}\right)}
\newcommand\sbrackets[1]{\left[{#1}\right]}

\newcommand\abrackets[1]{\left\langle{#1}\right\rangle}
\newcommand\divergence[2][\rbrackets]{\nabla \cdot #1{#2}}
\newcommand\curl[2][\rbrackets]{\nabla \times #1{#2}}

\newcommand\kbol{k_{\rm B}}

\newcommand\vel{\mathbf{v}}

\newcommand\xhat{\hat{\mathbf{x}} }
\newcommand\yhat{\hat{\mathbf{y}} }
\newcommand\zhat{\hat{\mathbf{z}} }

\newcommand\kpeak{k_{\rm pk}}
\newcommand\lpeak{\lambda_{\rm pk}}
\newcommand\Eturb{E_{\rm turb}}
\newcommand\Ecomp{\mathcal{E}_{\rm comp}}
\newcommand\Esh{\mathcal{E}_{\rm sh}}
\newcommand\PS{\mathcal{P}}
\newcommand\Eperp{E_{\rm K,perp}}
\newcommand\Epara{E_{\rm K,para}}
\newcommand\Ekin{E_{\rm K}}
\newcommand\Emag{\delta E_{\rm B}}
\newcommand\tcross{t_{\rm cr}}
\newcommand\tdec{t_{\rm dec}}
\newcommand\tdeck{t_{\rm dec}^K}

\newcommand\kphat{\mathbf{\hat{k}}_R}
\newcommand\bhat{\hat{\mathbf{b}}}
\newcommand\Alfven{Alfv\'{e}n }
\newcommand\Alfvenic{Alfv\'{e}nic }
\newcommand\muG{\;\mu{\rm G}}

\shorttitle{Decaying MHD Turbulence in the Multiphase ISM}
\shortauthors{Kim \& Basu}
\begin{document}

\title{Long-Term Evolution of Decaying MHD Turbulence in the Multiphase ISM}

\author{Chang-Goo Kim and Shantanu Basu}

\affil{Department of Physics \& Astronomy, University of Western Ontario,
London, Ontario N6A 3K7, Canada}
\email{ckim256@uwo.ca, basu@uwo.ca}
%\slugcomment{Last Modified: \today}

\begin{abstract}
Supersonic turbulence in the interstellar medium (ISM) is believed to decay
rapidly within a flow crossing time irrespective of the degree of
magnetization.  However, this general consensus of decaying magnetohydrodynamic
(MHD) turbulence relies on local isothermal simulations, which are unable to
take into account the roles of global structures of magnetic fields and the
ISM. Utilizing three-dimensional MHD simulations including interstellar cooling
and heating, we investigate decaying MHD turbulence within cold neutral medium
sheets embedded in warm neutral medium.  Early evolution of turbulent kinetic
energy is consistent with previous results for decaying compressible MHD
turbulence characterized by rapid energy decay with a power-law form of
$E\propto t^{-1}$ and by short decay time compared to a flow crossing time. If
initial magnetic fields are strong and perpendicular to the sheet, however,
long-term evolution of the kinetic energy shows that a significant amount of
turbulent energy ($\sim 0.2E_0$) still remains even after ten flow crossing
times for models with periodic boundary conditions.  The decay rate is also
greatly reduced as the field strength increases for such initial and boundary
conditions, but not if the boundary conditions are that for a completely
isolated sheet. We analyze velocity power spectra of the remaining turbulence
to show that in-plane, incompressible motions parallel to the sheet dominate at
later times.  
\end{abstract}


\section{Introduction\label{sec:intro}}

Supersonic turbulence is ubiquitous in the interstellar medium (ISM) and is a
key ingredient of star formation processes (see \citealt{elm04,mac04,mck07} for
recent reviews). The characteristics and properties of turbulence have long
been studied since the pioneering study for incompressible, hydrodynamic
turbulence by \citet{kol41}, which is applicable mostly to terrestrial flows.
For incompressible turbulence, turbulent energy is transferred from larger to
smaller scales via eddy cascading all the way down to the dissipation scale,
where viscosity is effective. Turbulence in the ISM, however, is expected to
depart largely from this classical picture mostly due to its compressibility
with supersonic Mach number and non-negligible degree of magnetization. A
general theory of incompressible magnetohydrodynamic (MHD) turbulence has been
developed by \citet{gol95} and tested by numerical simulations for both
incompressible and compressible gas flows (e.g.,
\citealt{cho00a,mar01b,lit01,cho03,ves03}). Similar energy cascading processes
occur in MHD turbulence as well, while the eddies are elongated along the
direction of the magnetic field.  

The time evolution of turbulent energy is expected to follow a power-law form
of $E\propto t^{-\eta}$. The exponent $\eta$ of incompressible hydrodynamic
turbulence has been analytically suggested to be $10/7$ from Kolmogorov theory
\citep{kol41} and experimentally measured to be in a range from $1.2$ to $2$
\citep{1993PhRvL..71.2583S}.  Since supersonic gas flows naturally induce shock
waves, that provide additional energy dissipation, turbulent energy is expected
to be dissipated away rapidly in less than than lifetime of a molecular cloud
in the absence of driving. This theoretical expectation contradicts with
universally observed suprathermal linewidths within molecular clouds (e.g.,
\citealt{lar81}).  Magnetic fields have then come into the spotlight to resolve
this contradiction. \citet{aro75} have suggested non-dissipative linear \Alfven
waves as sub-\Alfvenic turbulence to explain the observed turbulence without
the need for external energy driving (see also \citealt{zwe83,elm85}).

Recently, numerical simulations allowed to explore the nonlinear evolution of
supersonic MHD turbulence and revealed that the previous belief of long-lived
sub-\Alfvenic turbulence is incorrect. \citet{gam96b} have shown that kinetic
energy dissipates quickly in one-dimensional MHD simulations of isothermal
compressible decaying turbulence. This work has been extended to
three-dimensional local periodic-box simulations by several authors (e.g.,
\citealt{mac98,sto98,pad99a}). The results from simulations for compressible
MHD turbulence have converged to a rapid decay of turbulence in the ISM, as
$E\propto t^{-1}$, regardless of magnetic field strengths. If the energy
transfer to larger scales through the inverse cascade is prevented by energy
injection on the scale of the system, \citet{cho03} have suggested $\eta$ would
become closer to 2. Supersonic turbulence in molecular cloud conditions loses
half of the initial energy within a flow crossing time \citep{ost01}, usually
discuss comparable to a free-fall time of the cloud.  Based on these numerical
studies, the universal supersonic turbulence observed in clouds is now believed
to arise from inexhaustible energy input mechanisms, e.g., stellar feedback,
gas-dynamical instabilities, galactic rotation, and so on (see
\citealt{mac04,elm04} and references therein). 

Although a general consensus of rapidly decaying compressible MHD
turbulence has been achieved during the last two decades, the conclusion relies
heavily on the results of local isothermal simulations (see also local models
for the multiphase ISM; \citealt{kri02,kri04}). Since these models represent a
local patch of a homogeneous medium embedded within a larger cloud, it is not
directly applicable to turbulence in the diffuse multiphase ISM. Thermal
processes in the ISM naturally induce an inhomogeneity, giving rise to two
distinct atomic phases, a cold neutral medium (CNM) and a warm neutral medium
(WNM), whose density and temperature differ by about two orders of magnitude
\citep{fie69, wol95}. High-resolution 21 cm absorption line observations using
the Arecibo telescope have shown that the CNM is generally observed within
sheets rather than isotropic clouds \citep{hei03}.  These CNM sheets have
magnetic fields that dominate the thermal pressure of the ISM, with a median
strength of $6\muG$ \citep{hei05b,cru12}, while the orientation of the magnetic
fields to the sheet is somewhat uncertain. Such commonly observed sheet-like
structures of the CNM with strong magnetic fields threading both CNM and WNM
suggest that idealized local isothermal models have clear limitations in
application to turbulence of the diffuse multiphase ISM. The global structures
of the ISM and magnetic fields should be taken into account to study ISM
turbulence properly. 

\citet{kud03,kud06} have considered a stratified molecular cloud using a
Lagrangian fluid description for the temperature to model a cold cloud within a
hot medium. Their 1.5 dimensional simulations show lower dissipation rates than
local isothermal simulations, but it is unclear whether this results from a
lack of dimensionality \citep{mac98,ost99} or the global structure
they considered.  Using two-dimensional simulations of thin sheets with the external
magnetic field calculated as a potential field, \citet{bas10} have shown that
global magnetic tension driven modes produce long-lasting turbulence in the
flux-freezing limit, in contrast to previous local models. Without a thin-disk
approximation, however, three-dimensional stratified cloud models with
Lagrangian fluid elements have reported that energy dissipation occurs
differently from \citet{bas10}, but about 10\% of the initial kinetic
energy remains at the final stage \citep{kud11}. 

Motivated by limitations of local isothermal simulations and recent simulations
for thin sheets, we investigate the long-term evolution of decaying turbulence.
We utilize full three-dimensional simulations with interstellar cooling and
heating in order to understand turbulence within the diffuse multiphase ISM.  Our
numerical models provide alternative perspectives on decaying turbulence within
CNM sheets, including the effect of magnetic fields anchored in the ambient WNM as
well as non-isothermality and field orientations. This paper is organized as
follows. In Section~\ref{sec:method}, we describe numerical methods and initial model
setups.  In Section~\ref{sec:turb_evol}, we present our main results using temporal
evolution of turbulent kinetic energy for models with different magnetic field
strengths and orientations.  In Section~\ref{sec:turb_chr}, we analyze turbulence
characteristics in detail, especially strong field models.  In
Section~\ref{sec:discussion}, we discuss our model setup and outcomes in
comparison with observations, and Section~\ref{sec:summary} summarizes our main
results.

\section{Numerical Methods \& Models\label{sec:method}}

In order to investigate the decaying MHD turbulence in the multiphase ISM, we
solve a set of ideal MHD equations with cooling, heating, and thermal
conduction: 
\begin{equation}\label{eq:cont}
\pderiv{\rho}{t}+\divergence{\rho\vel}=0,
\end{equation}
\begin{equation}\label{eq:mom}
\pderiv{\rho\vel}{t}+
\divergence[\sbrackets]{\rho\vel\vel
-\frac{\mathbf{B}\mathbf{B}}{4\pi}+\rbrackets{P+\frac{B^2}{8\pi}}\mathbf{I}}=0,
\end{equation}
\begin{equation}\label{eq:energy}
\pderiv{E}{t}+\divergence[\sbrackets]{\rbrackets{E+P+\frac{B^2}{8\pi}}\vel
-\frac{\mathbf{B}}{4\pi}(\mathbf{B}\cdot\vel)-\mathcal{K}\nabla T}= -\rho\mathcal{L},
\end{equation}
\begin{equation}\label{eq:ind}
\pderiv{\mathbf{B}}{t}-\curl{\vel\times\mathbf{B}}=0,
\end{equation}
\begin{equation}\label{eq:div}
\nabla\cdot\mathbf{B}=0,
\end{equation}
where $\rho$, $\mathbf{v}$, P, and $\mathbf{B}$ are the total gas mass density,
velocity vector, gas thermal pressure, and magnetic field, respectively, as
usual. $E$ is the total energy density 
\begin{equation}
E=\frac{P}{\gamma-1}+\frac{\rho v^2}{2}+\frac{B^2}{8\pi},
\end{equation}
where $v^2=\mathbf{v}\cdot\mathbf{v}$ and $B^2=\mathbf{B}\cdot\mathbf{B}$.  We
adopt the ideal gas law $P=(\gamma-1)e$ with $\gamma=5/3$, where $e$ is the
internal energy density.  The thermal pressure is $P=1.1n\kbol T$ for 10\% of
Helium abundance.  Here $n$ is the number density of hydrogen, and hence
$\rho=1.4\mh n$.  The thermal conductivity is
$\mathcal{K}=10^5{\rm\;erg\;s^{-1}\;cm^{-1}\;K^{-1}}$. The source term in the
energy equation (Eq.~(\ref{eq:energy})) is the net volumetric cooling given by
$\rho\mathcal{L}\equiv n[n\Lambda(T)-\Gamma]$.  Following \citet{koy02}, we
adopt the simple fitting formula for dominant interstellar cooling, mainly due
to [\ion{C}{2}], [\ion{O}{1}] fine structure and Ly$\alpha$ recombination
lines, and the constant heating rate due to the photoelectric effect on grains
by FUV photons (\citealt{kko08} for a typo corrected version):
\begin{equation}\label{eq:cool}
 \Lambda(T)=2\times 10^{-19}\exp\left(\frac{-1.184\times10^5}{T+1000}\right)
 +2.8\times10^{-28}\sqrt{T}\exp\left(\frac{-92}{T}\right)
{\rm \;erg} \cm^3 \;{\rm s^{-1}},
\end{equation}
\begin{equation}\label{eq:heat}
\Gamma=2\times10^{-26}\ergs.
\end{equation}

We solve the governing equations using the {\it Athena} code that employs the
high order Godunov method \citep{sto08,sto09}.  Among the various techniques
provided by the {\it Athena} code to solve the MHD equations, we adopt van Leer
integrator \citep{sto09} with Roe's Riemann solver and second-order spatial
reconstruction scheme. {\it Athena} satisfies the divergence-free condition
within machine precision via the constrained transport algorithm \citep{eva88}.
For the cooling source term, we solve it separately using a fully implicit
method with a Newton-Raphson iteration \citep{pio04,kko08} and update
temperature before entering the integrator steps for conservation equations.
To ensure the code stability, we reduce the time step until the temperature
change is smaller than 50\% of the previous temperature.  An isotropic thermal
conduction is solved explicitly as well. 

For an initial state, we assume a thin CNM sheet in a thermal pressure
equilibrium with the surrounding WNM. So we first solve thermal equilibrium
equation, $\Gamma=n\Lambda(T)$, for a given equilibrium pressure $P_0$ to find
the density and temperature of the CNM and WNM, $(n_c, T_c)$ and $(n_w, T_w)$,
respectively.  For our fiducial parameter of initial thermal pressure,
$P_0/\kbol=3000\Punit$ \citep{wol03,jen11}, the equilibrium density and
temperature are $(48\pcc,57\Kel)$ and $(0.4\pcc, 6500\Kel)$ for the CNM and
WNM, respectively.  We then assign a uniform CNM for $|z|<\Delta_c/2$, where
$\Delta_c=2\pc$ is the thickness of the CNM sheet, and the WNM fills other
regions of our cubical box with side lengths of $L=20\pc$.  The uniform
magnetic field is assigned with the field strength of
\begin{equation}\label{eq:B0}
B_0=3.2\;\beta^{-1/2}\rbrackets{\frac{P_0/\kbol}{3000\Punit}}^{1/2}{\rm \;\mu G},
\end{equation}
where $\beta\equiv 8\pi P_0/B_0^2$ is the plasma beta that parameterizes
the magnetic field strength using the ratio of thermal to magnetic pressure.

We assign initial velocity perturbations using independent realizations of a
Gaussian random field with a power spectrum (PS) $|v_k^2|\propto
k^6\exp(-8k/\kpeak)$ for each velocity component. Here, $\kpeak=2\pi/\lpeak$ is
the wavenumber at the peak of the PS. We adopt $\kpeak L/(2\pi)=8$ and the
amplitude of the velocity field $\sigma_{\rm 1D,0}=2\kms$.  The flow crossing
time at the scale $\lpeak=L/8$ is then $\tcross\equiv\lpeak/\sigma_{\rm
1D,0}=1.2\Myr$, which will be considered as a time unit throughout the paper.
This represents a reference time scale for the initial turbulent decay and can
be used for comparison with previous work. Since the CNM has a sound speed of
$c_c\equiv(P_0/\rho_c)^{1/2}=0.6\kms$ and an \Alfven speed of $v_{A,c}\equiv
B_0/(4\pi\rho_c)^{1/2}=(2/\beta)^{1/2}c_c$, our choice of $\sigma_{\rm 1D,0}=2\kms$
is supersonic and sub-\Alfvenic for $\beta=0.1$, slightly super-\Alfvenic for
$\beta=1$, and super-\Alfvenic for $\beta=10$ in the CNM sheet (see
Figure~\ref{fig:Mpdf} for details of Mach number distribution).  On the other
hand, the sound speed of the WNM is $c_w=6.6\kms$ so that initial turbulence
for the WNM is always subsonic and sub-Alfv\'enic. The sound and \Alfven wave
crossing times in the vertical direction are
$t_s\equiv(L-\Delta_c)/c_w+\Delta_c/c_c\sim5\tcross$ and
$t_A=t_s(\beta/2)^{1/2}$, respectively.  The \Alfven wave crossing times yield
$t_A/\tcross=1.1$, 3.4, and 11 for $\beta=0.1$, 1, and 10, respectively.

It is noteworthy that the saturated state of turbulence is achieved after 2 or
3 largest-eddy turnover times (e.g., \citealt{sto98}). Thus, the early
evolution till (2-3)$\tcross$ in our models may reflect transient features
associated with a particular choice of initial conditions. The scaling
relationships between fluid properties should be addressed after the saturation
of turbulence. However, the slope of the time evolution of turbulent
kinetic energy remains similar to that when starting from a saturated state of
turbulence (compare \citealt{mac98} with \citealt{sto98}).

The simulation parameters are summarized in Table~\ref{tbl:model}.  Here we
have three different model series with various magnetic field configurations:
Series A has vertical magnetic fields, $\mathbf{B}=B_0\zhat$, and anisotropic
turbulence (no vertical motions, $\sigma_{\rm z,0}=0\kms$); Series B has
horizontal magnetic fields, $\mathbf{B}=B_0\xhat$, and anisotropic turbulence;
and Series C has vertical magnetic fields, $\mathbf{B}=B_0\zhat$, and isotropic
turbulence (non-zero vertical motions, $\sigma_{\rm z,0}=2\kms$).  Each series
has additional labels of S, I, W, and H, corresponding to strong ($\beta=0.1$),
intermediate ($\beta=1$), weak ($\beta=10$) magnetic fields, and purely
hydrodynamic ($\beta=\infty$), respectively. Note that the AH and BH models are
identical.

Boundary conditions (BCs) are periodic in the horizontal directions ($\xhat$
and $\yhat$). In the vertical direction, we adopt periodic boundary conditions
for quasi-global cases, representing regularly placed CNM sheets within larger
\ion{H}{1} clouds. We also run Series A with continuous BCs for a single,
isolated CNM sheet in order to study the effect of BCs (see
Section~\ref{sec:turb_evol_BC}).

The major diagnostics of our simulations are turbulent energies defined by
\begin{equation}\label{eq:keturb}
\Ekin =\frac{1}{2} \int \rho v^2 dV
\end{equation}
and
\begin{equation}\label{eq:meturb}
\Emag =\frac{1}{8\pi}\int (B^2 - B_0^2)dV.
\end{equation}
The total turbulent energy can be obtained by $\Eturb=\Ekin+\Emag$.  Note that,
by definition, the kinetic and magnetic energies are weighted by mass and
volume, respectively. Since total mass is dominated by the CNM, and the total
turbulent energy is dominated by the kinetic energy except in the very early
amplification phase ($<\tcross$) of the turbulent magnetic energy for models
with $\beta=0.1$, $\Ekin$ is the best probe of turbulent energy within the CNM.

We carry out a convergence test for the AI model using numbers of grid zones
from $32^3$ to $512^3$. We confirmed that both turbulent kinetic and magnetic
energies converge from $256^3$ upward, so we adopt $256^3$ grid zones for our
standard numerical resolution for all simulations, which gives a grid size of
$\Delta=0.08\pc$.

For a quick overview of our results, we list the decay times of the total and
kinetic turbulent energy, $\tdec$ and $\tdeck$ (defined by the time when the
initial total and kinetic energy is reduced by 50\%), respectively. We also
list the remaining kinetic energy at $\tcross$ and $10\tcross$ as well as the
ratio of turbulent magnetic to kinetic energy. Note that we use only the
kinetic energy in the $\xhat$-$\yhat$ plane (XY plane hereafter) for Series C
to avoid contamination from an initial vertical expansion (see
Section~\ref{sec:turb_evol_IC} for details).  These values can be compared to
previous results (e.g., \citealt{ost01}).

\begin{deluxetable}{lccccccccc}
% Model Parameters
\tabletypesize{\footnotesize} \tablewidth{0pt} 
\tablecaption{Model Parameters \& Simulation Results\label{tbl:model}} 
\tablehead{ 
\colhead{Model} &
\colhead{$\beta$} & 
\colhead{$\bhat$} &
\colhead{$\sigma_{z,0}$\tablenotemark{a}} &
\colhead{$\tdec$\tablenotemark{b,c}} &
\colhead{$\tdeck$\tablenotemark{b,c}} &
\colhead{$\Ekin(\tcross)$\tablenotemark{b,d}} &
\colhead{$\Ekin(10\tcross)$\tablenotemark{b,d}} &
\colhead{$\Emag/\Ekin(\tcross)$} &
\colhead{$\Emag/\Ekin(10\tcross)$}
}
\startdata 
AS & 0.1      &  $\zhat$ & 0 & 0.65 & 0.44 & 0.32 & 0.18 & 0.19 & 0.07\\
AI & 1        &  $\zhat$ & 0 & 0.40 & 0.27 & 0.21 & 0.06 & 0.26 & 0.12\\
AW & 10       &  $\zhat$ & 0 & 0.31 & 0.28 & 0.18 & 0.02 & 0.15 & 0.40\\
AH\tablenotemark{e} & $\infty$ & \nodata & 0 & 0.29 & 0.29 & 0.20 & 0.03 & \nodata & \nodata\\
BS & 0.1      &  $\xhat$ & 0 & 0.37 & 0.18 & 0.15 & 0.02 & 0.43 & 0.07\\ 
BI & 1        &  $\xhat$ & 0 & 0.32 & 0.22 & 0.13 & 0.01 & 0.61 & 0.48\\
BW & 10       &  $\xhat$ & 0 & 0.30 & 0.26 & 0.15 & 0.02 & 0.32 & 0.86\\
CS & 0.1      &  $\zhat$ & 2 & 0.49 & 0.11 & 0.19 & 0.06 & 0.22 & 0.01\\
CI & 1        &  $\zhat$ & 2 & 0.40 & 0.21 & 0.14 & 0.02 & 0.32 & 0.08\\
CW & 10       &  $\zhat$ & 2 & 0.36 & 0.27 & 0.17 & 0.01 & 0.21 & 0.45\\
CH & $\infty$ &  \nodata & 2 & 0.33 & 0.33 & 0.22 & 0.02 & \nodata & \nodata \\
\enddata
\tablenotetext{a}{In units of $\kms$}
\tablenotetext{b}{Only horizontal components ($\xhat$ and $\yhat$) of turbulent
kinetic energy ($\Eperp$) are used for Series C to exclude the effect of
initial vertical expansion. See \S\ref{sec:turb_evol_IC} for details.} 
\tablenotetext{c}{In units of $\tcross$}
\tablenotetext{d}{In units of $\Ekin(0)$}
\tablenotetext{e}{The AH model is also referred to as the BH model in the text
since they are identical.}
\end{deluxetable}

\section{Results\label{sec:result}}

Since the main purpose of this paper is to investigate the long-term evolution
of the turbulent energy and its characteristics in the multiphase ISM, we will
begin by describing the temporal evolution of turbulent kinetic energy for each
model.  Then we will analyse the characteristics of the remaining turbulence of
each model, putting particular emphasis on models with strong magnetic fields
($\beta=0.1$). These models have a significant amount of turbulent energy at
later times.

\subsection{Evolution of Turbulence Energy\label{sec:turb_evol}}

\subsubsection{Effect of Initial Conditions\label{sec:turb_evol_IC}}

\begin{figure*}
\plotone{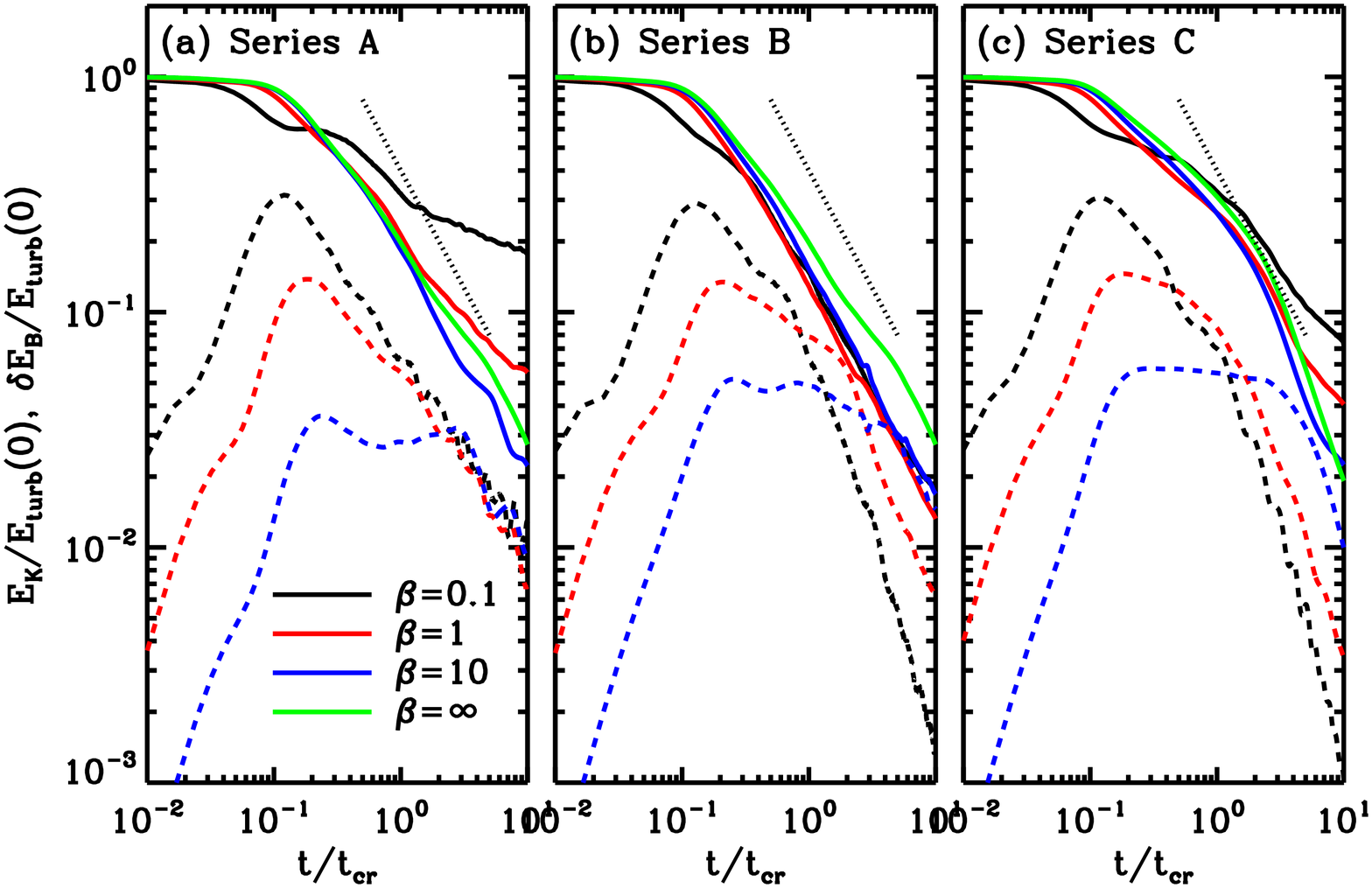}
\caption{Time evolution of turbulent kinetic and magnetic energy, $\Ekin$
(\emph{solid}) and $\Emag$ (\emph{dashed}), respectively, normalized by the
initial turbulent energy, $\Eturb(0)$.  (a) Series A: initial magnetic fields
are vertical ($\bhat=\zhat$), and initial velocity fields are anisotropic (no
vertical motions); (b) Series B: initial magnetic fields are horizontal
($\bhat=\xhat$), and initial velocity fields are anisotropic; and (c) Series C:
initial magnetic fields are vertical ($\bhat=\zhat$), and initial velocity
fields are isotropic in all three dimensions.  The dotted line in each panel
denotes a reference line for a power-law form of $E\propto t^{-1}$.
\label{fig:tevol}}
\end{figure*}

Figure~\ref{fig:tevol} plots $\Ekin/\Eturb(0)$ (\emph{solid}) and
$\Emag/\Eturb(0)$ (\emph{dashed})  as a function of time, $t/\tcross$, for all
models with periodic BCs for all directions.  In panels (a) and (b), we show
models in Series A and B, respectively, whose initial velocity perturbations
are anisotropic without initial vertical motions ($\sigma_{z,0}=0$). Note that
hydrodynamics models (AH and BH models; \emph{green}) in panels (a) and (b) are
identical.  Series C is shown in panel (c).  Initial velocity perturbations
deform magnetic field lines, and the kinetic energy is partially converted to
the turbulent magnetic energy.  The perturbed magnetic energy reaches its
maximum at around 0.1-0.2$\tcross$, having 5-30\% of $\Eturb(0)$. The
peak magnetic energy increases with increasing initial field strength, and the
time to reach the peak energy is slightly delayed as initial field strength
decreases. After the time when $\Emag$ is maximal, both turbulent kinetic and
magnetic energies begin to decay.  Although the turbulent kinetic (and
magnetic) energy in all models decays rapidly and loses about half of the
initial energy within one crossing time as in previous isothermal simulations
\citep[e.g.,][]{mac98,sto98,pad99a,ost01}, the late time evolution is different
from model to model as magnetic field strengths vary.

The most notable result is shown in Series A (Figure~\ref{fig:tevol}(a)); an
increasing amount of turbulent energy remains as magnetic fields get stronger.
The early evolution up to one or two crossing times is similar to previous
isothermal models.  There is no significant difference depending on magnetic
field strengths. The decay times, defined as the time when the initial energy
is reduced by $50\%$, are $\tdec/\tcross=0.65$, $0.40$, $0.31$, and $0.29$, for
$\beta = 0.1$, $1$, $10$, and $\infty$, respectively (see
Table~\ref{tbl:model}). For kinetic energy alone, we obtain
$\tdeck/\tcross=0.44$, $0.27$, and $0.28$, for $\beta=0.1$, $1$, and $10$,
respectively. These values are roughly consistent with the results of
\citet{ost01}, despite different initial velocity perturbations, gas
structures, equation of state, and magnetic field configurations.  The energy
decay follows a power-law type evolution with $\Eturb\propto t^{-1}$ as in
\citet{mac98} and \citet{sto98}.  However, the energy decay becomes shallower
after a few crossing times depending on the strength of the magnetic field. For
the AS model, it is clearly seen that a significant amount of turbulent energy
remains even at $10\tcross$ ($\Ekin/\Eturb(0)=0.18$ at $t/\tcross=10$).

With horizontal magnetic fields (Series B; Figure~\ref{fig:tevol}(b)), in
contrast to Series A, the magnetic field strength does not affect the evolution
of the turbulent kinetic energy. The kinetic energy keeps decaying roughly with
a power-law form of $\Eturb\propto t^{-1}$.  The decay times for the total
energy are slightly longer for models with stronger magnetic fields, while the
trend is reversed for kinetic energy (see Table~\ref{tbl:model}). This is also
seen in \citet{ost01} due to the initial amplification of magnetic fields. The
turbulent kinetic energy remains only about 15\% at $t/\tcross=1$ and less than
2\% at $t/\tcross=10$ for all $\beta$.

Note that \citet{ost99} have also shown a shallower energy decay for stronger
magnetic field strength in their 2.5D simulations with horizontal magnetic
fields. As they argued, this is likely due to the lack of dissipation in the
symmetric direction.\footnote{We have run similar simulations to \citet{ost99}
and found that the shallower decay of kinetic energy is indeed due to a
shallower decay of the vertical energy component. The horizontal components are
still decaying as $E\propto t^{-1}$.} Since here we allow the vertical degree
of freedom, this shallower decay cannot be seen.  Rather, we can see the effect
of magnetic fields when the field is vertical (Series A), which is not because
of the numerically suppressed dissipation in one direction but because of the
surviving incompressible horizontal motions in such magnetic field
configurations and periodic BCs (see Section~\ref{sec:turb_evol_BC} for more
details).

In Figure~\ref{fig:tevol}(c), models with isotropic velocity perturbations
(Series C) show different evolutionary behavior.  It is mainly due to the
initial vertical expansion, arising from an imbalance of initial turbulent
pressure. The initial CNM sheet is in thermal pressure equilibrium with the
surrounding WNM. However, vertical turbulent motions give rise to an additional
turbulent pressure in the vertical direction, which is greater in the CNM
because of its greater density.  A shallower decay of kinetic energy is
observed until the vertical expansion ends at about the half-thickness crossing
time $(L/2)/\sigma_{\rm 1D,0}=4\tcross$.  Since the kinetic energy evolution is
contaminated by the shallower decay of the vertical energy component (similar
to \citet{ost99} but for a different reason), it is necessary to isolate the
effect of the initial vertical expansion in order to understand turbulent
energy evolution correctly.

\begin{figure}
\plotone{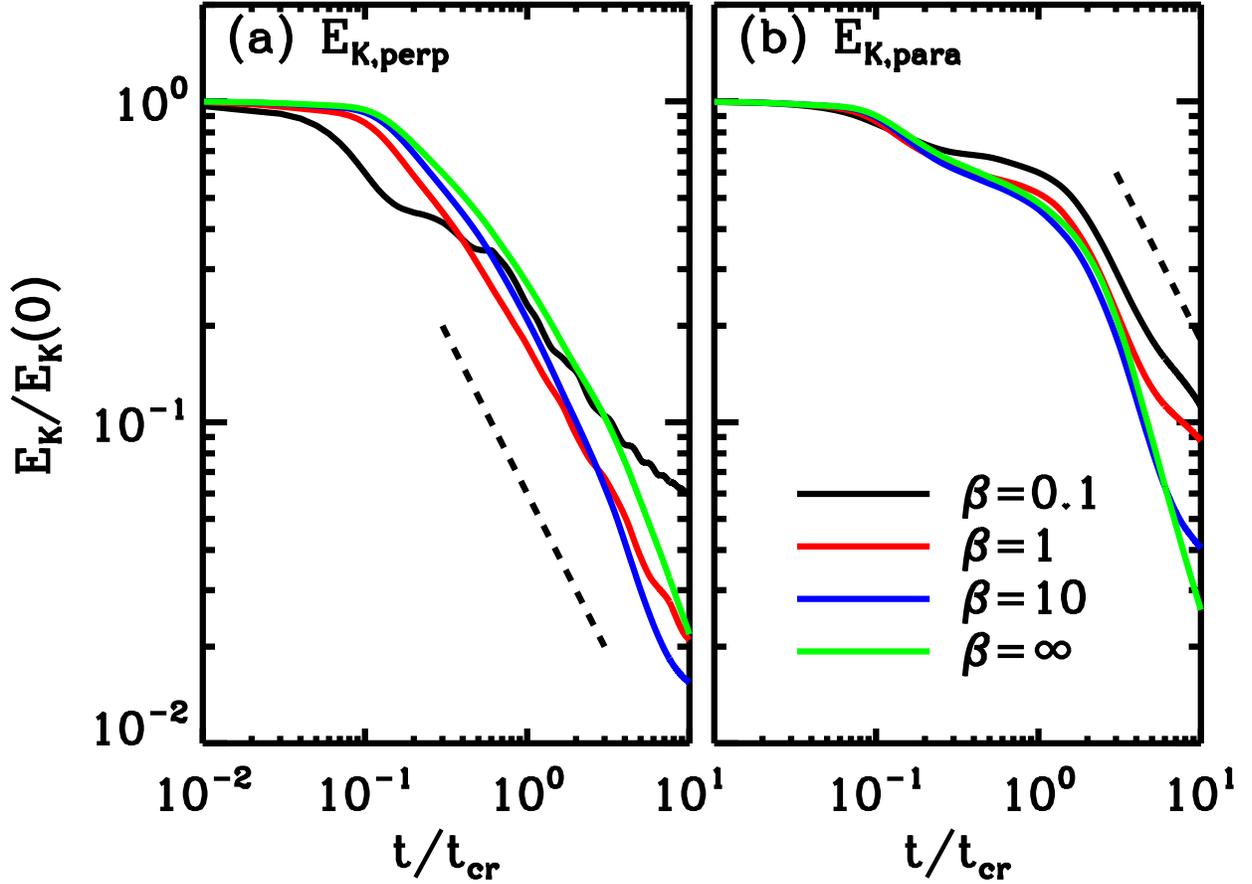}
\caption{Time evolution of turbulent kinetic energy in the directions (a)
perpendicular, $\Eperp/\Eperp(0)$, and (b) parallel, $\Epara/\Epara(0)$, to the
initial magnetic field for Series C.  Note that $\Eperp=E_{\rm K,x}+E_{\rm
K,y}$ is two times greater than $\Epara=E_{\rm K,z}$ at the initial time.  The
dashed line in each panel denotes a reference line for a power-law form of
$\Ekin\propto t^{-1}$.  \label{fig:tevol_C}}
\end{figure}

We separate the turbulent kinetic energy into perpendicular ($\Eperp=E_{\rm
K,x}+E_{\rm K,y}$) and parallel ($\Epara=E_{\rm K,z}$) components to the
magnetic field direction.  Figure~\ref{fig:tevol_C} plots  (a)
$\Eperp/\Eperp(0)$ and (b) $\Epara/\Epara(0)$ as a function of time for Series
C.  Note that $\Eperp$ is two times greater than $\Epara$ initially since the
initial velocity field is isotropic in Series C.  As shown in panel (b), the
parallel component of the kinetic energy is hardly decaying until
$t\sim2\tcross$. Since the surrounding WNM has a small density, the CNM expands
nearly freely in the vertical direction with less dissipation. After this
expansion, the parallel component decays as rapidly as a power-law of $t^{-1}$
irrespective of the field strengths.

However, the perpendicular component of kinetic energy decays similarly to that
in Series A; a power-law decay with $\Ekin\propto t^{-1}$ at an early stage and
a shallower decay at later times.  The strong field model (CS model;
\emph{black}) contains about an order of magnitude more energy at the end of
the simulation compared to the CW and CH models, while the remaining energy is
significantly dissipated already: at $t/\tcross=10$, $\Eperp/\Eperp(0)=5.6\%$,
$1.8\%$, $1.4\%$, and $1.6\%$ for the CS, CI, CW, and CH models, respectively.

The evolution of the perturbed magnetic energy also shows a significant
difference between the model Series.  In Series A, the fluid motions are
assigned only in the perpendicular plane to the initial magnetic field. Thus,
the perturbed magnetic energy over the whole evolutionary stage is dominated by
the parallel component, $\delta B_z^2$, due to converging and diverging fluid
motions.  The generation of the perpendicular component, $\delta B_x^2 + \delta
B_y^2$, is limited, especially for the strong field case, where magnetic field
lines are hardly bent. Only the AW model has comparable energy in each
component at later times.  However, in Series B and C, there are non-negligible
fluid motions in the direction parallel to the magnetic fields, which generates
the perpendicular component of the perturbed magnetic energy as well by
stretching and shrinking the field lines. Despite having a different amount of
energy stored in each component, the ratios of the total perturbed magnetic
energy to the initial turbulent energy at the peak are nearly the same. 
After (1-3)$\tcross$, when the fluid becomes really turbulent with proper
scaling relations, $\Emag$ decays rapidly in Series B and C as $\Emag\propto
t^{-2}$, similar to \citet{ost01}. On the other hand, since the remaining
kinetic energy is large enough to keep perturbing the magnetic field at later
times, the decay rate of $\Emag$ in Series A is reduced as well (the power-law
exponent is approximately $-1$). The models with weaker field strength in
Series B and C maintain their perturbed magnetic energy longer since the
magnetic field can be deformed even with weaker fluid motions.

\subsubsection{Effect of Boundary Conditions\label{sec:turb_evol_BC}}

\begin{figure}
\plotone{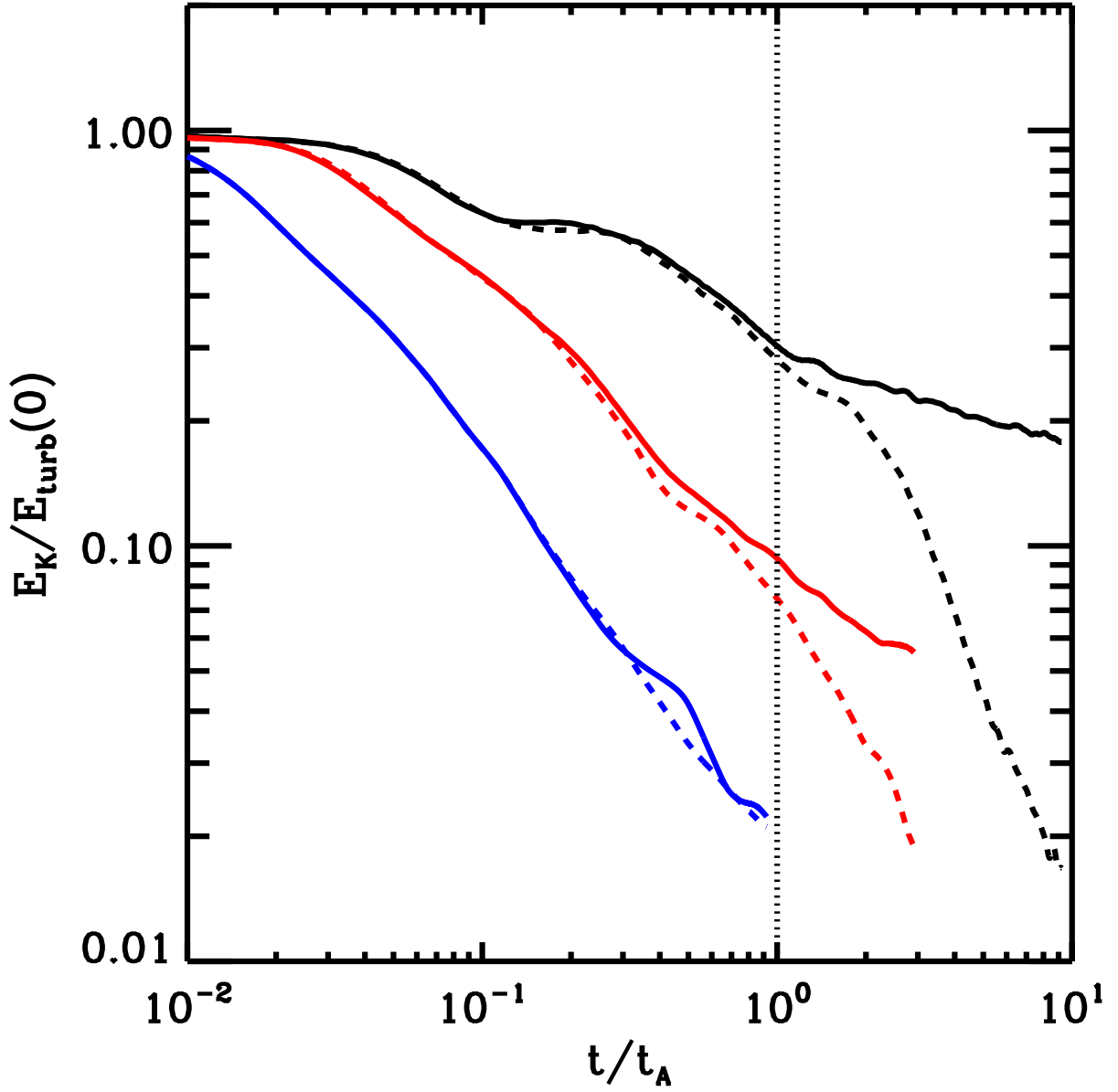}
\caption{ Evolution of turbulent kinetic energy, $\Ekin/\Eturb(0)$, as a
function of time in unit of the \Alfven wave crossing time,
$t_A=t_s(\beta/2)^{1/2}$ where $t_s\sim5\tcross$, for the AS (\emph{black}), AI
(\emph{red}), and AW (\emph{blue}) models.  The solid and dashed lines
respectively stand for periodic and continuous BCs at the vertical boundaries.
The vertical dotted line denotes $t=t_A$.  \label{fig:BCs}}
\end{figure}

The models with periodic BCs at the vertical boundaries presented in
Section~\ref{sec:turb_evol_IC} can be considered as a quasi-global case for CNM
sheets that are periodically placed within a large \ion{H}{1} cloud. Although
there might be many CNM sheets connected with magnetic fields in a \ion{H}{1}
cloud, the periodic BCs are still an oversimplified assumption. We thus
consider another extreme, a single isolated CNM sheet sandwiched by the WNM, by
running models with continuous BCs. In this case, waves can escape outward
freely. We only consider models in Series A with continuous BCs since Series B
is not much affected by vertical boundaries, and Series C is inappropriate to
run with continuous BCs due to the initial vertical expansion, which would
induce significant mass loss with continuous BCs.

Figure~\ref{fig:BCs} shows the time evolution of the normalized turbulent
kinetic energies $\Ekin/\Eturb(0)$ for the AS (\emph{black}), AI (\emph{red}),
and AW (\emph{blue}) models with periodic (\emph{solid}) and continuous
(\emph{dotted}) BCs at the vertical boundaries, $z=\pm L_z/2$.  Here, the time
is normalized by the \Alfven wave crossing time $t_A$, which differs for each
model. The kinetic energy evolution begins to differ at around $t=t_A$. After
one \Alfven wave crossing time, the overall evolution is greatly affected by
the BCs; the turbulent energy decays very rapidly with continuous BCs.

Since the initial velocity perturbations in Series A are assigned in the
perpendicular plane to the initial magnetic field, \Alfven waves are excited
initially and propagate toward the WNM along the field lines. They are not
dissipative themselves and hardly produce compressible waves (slow and fast
waves) since the mode coupling is in general very weak
\citep[e.g.,][]{cho02,cho03}.  It is evident in Figure~\ref{fig:BCs} that
travelling \Alfven waves that reenter to the CNM sheet after one \Alfven wave
crossing time are responsible for the reduced decay rates observed in the AS
and AI models.  Even with a strong magnetic field, the decay rate of turbulence
cannot be reduced for a single, isolated CNM sheet by storing energy in \Alfven
waves, as has long been considered since \citet{aro75}.  Rather, it seems that
maintaining and/or replenishing mechanisms for the \Alfven waves are necessary
to obtain a reduced decay rate.

\subsection{Turbulence Characteristics\label{sec:turb_chr}}

In Section~\ref{sec:turb_evol}, we find not only consistent results with
previous local isothermal simulations for early decay of turbulent kinetic
energy, but also new results of decaying turbulence with reduced rates in the
XY plane when the field is vertical and strong.  Since weak field models are
similar to hydrodynamic models and previous local simulations, here we put more
emphasis on strong field models for a detailed analysis of turbulence
characteristics.  Note that the intermediate field models also show similar
characteristics to the strong field models, but with less residual energy.

\subsubsection{AS Model\label{sec:AS}}

\begin{figure*}
\plotone{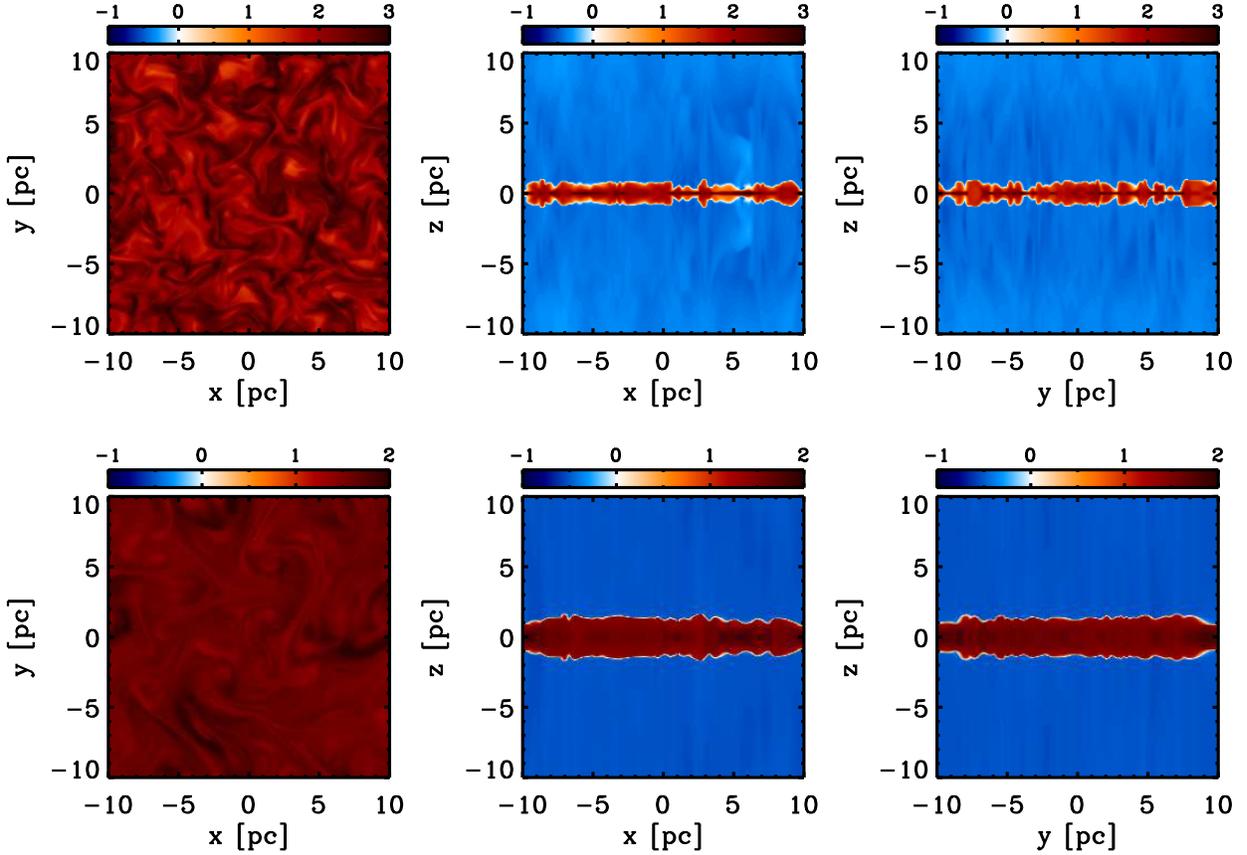}
\caption{Number density slices in logarithmic color scales at $z=0$
(\emph{left}), $y=0$ (\emph{middle}), and $x=0$ (\emph{right}) for the AS model at
$t/\tcross=1$ (\emph{top}) and $10$ (\emph{bottom}).  Note that we adjust the color
schemes in the color bars above each panel to associate a white color with the maximum
number density of the WNM ($n=1\pcc$) for a clearer distinction between the WNM and CNM
(\emph{plus} unstable phase).  Bluish and reddish colors denote the WNM and CNM
(\emph{plus} unstable phase), respectively.
\label{fig:snap_AS}} 
\end{figure*}

\begin{figure}
\plotone{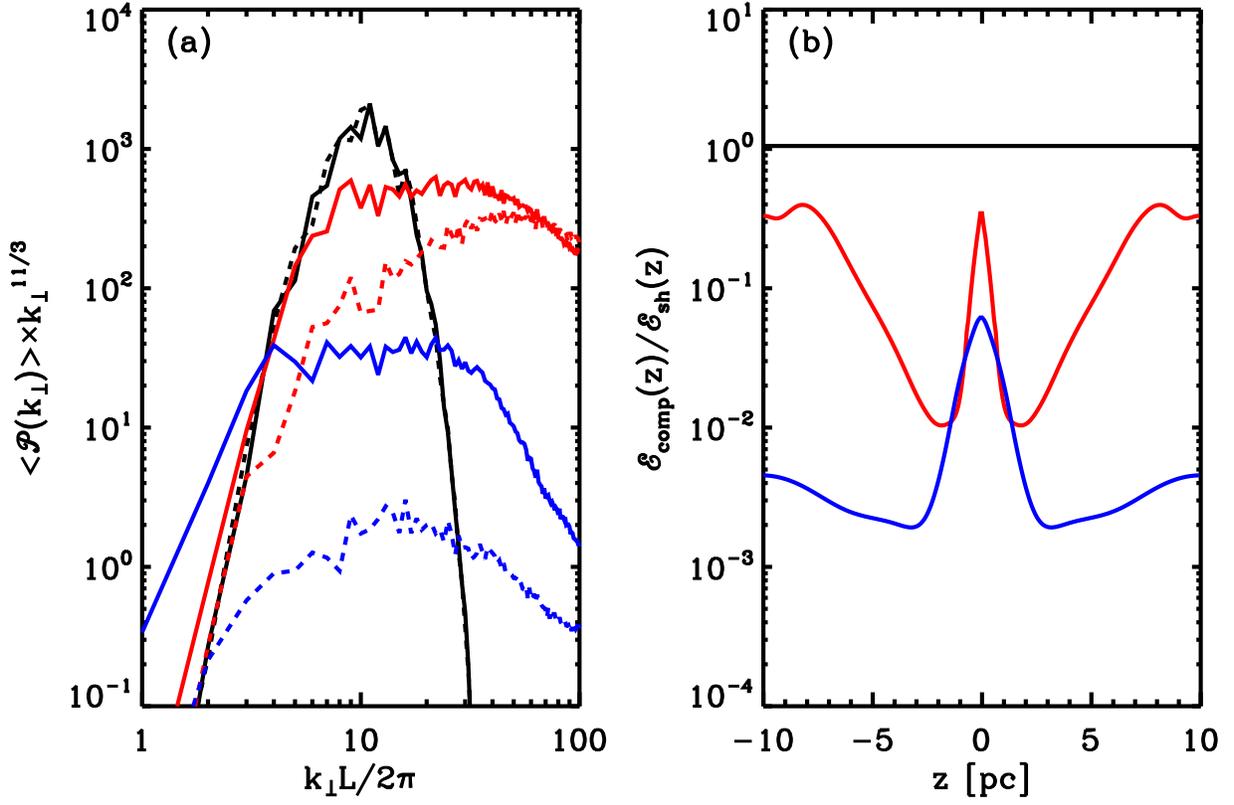}
\caption{(a) Cylindrically-binned, vertically-integrated PS of the AS model at
$t/\tcross=0$ (\emph{black}), $1$ (\emph{red}), and $10$ (\emph{blue}). The
solid and dashed lines stand for the shearing and compressible components,
respectively.  See Eq.(\ref{eq:Pk}) and text for details.  (b) Ratios between
the compressible and shearing components of the specific energy integrated over
$k_\perp$ as a function of $z$. See Eq.(\ref{eq:Ecomp}) and text for details.
\label{fig:power}}
\end{figure}

\begin{figure}
\plotone{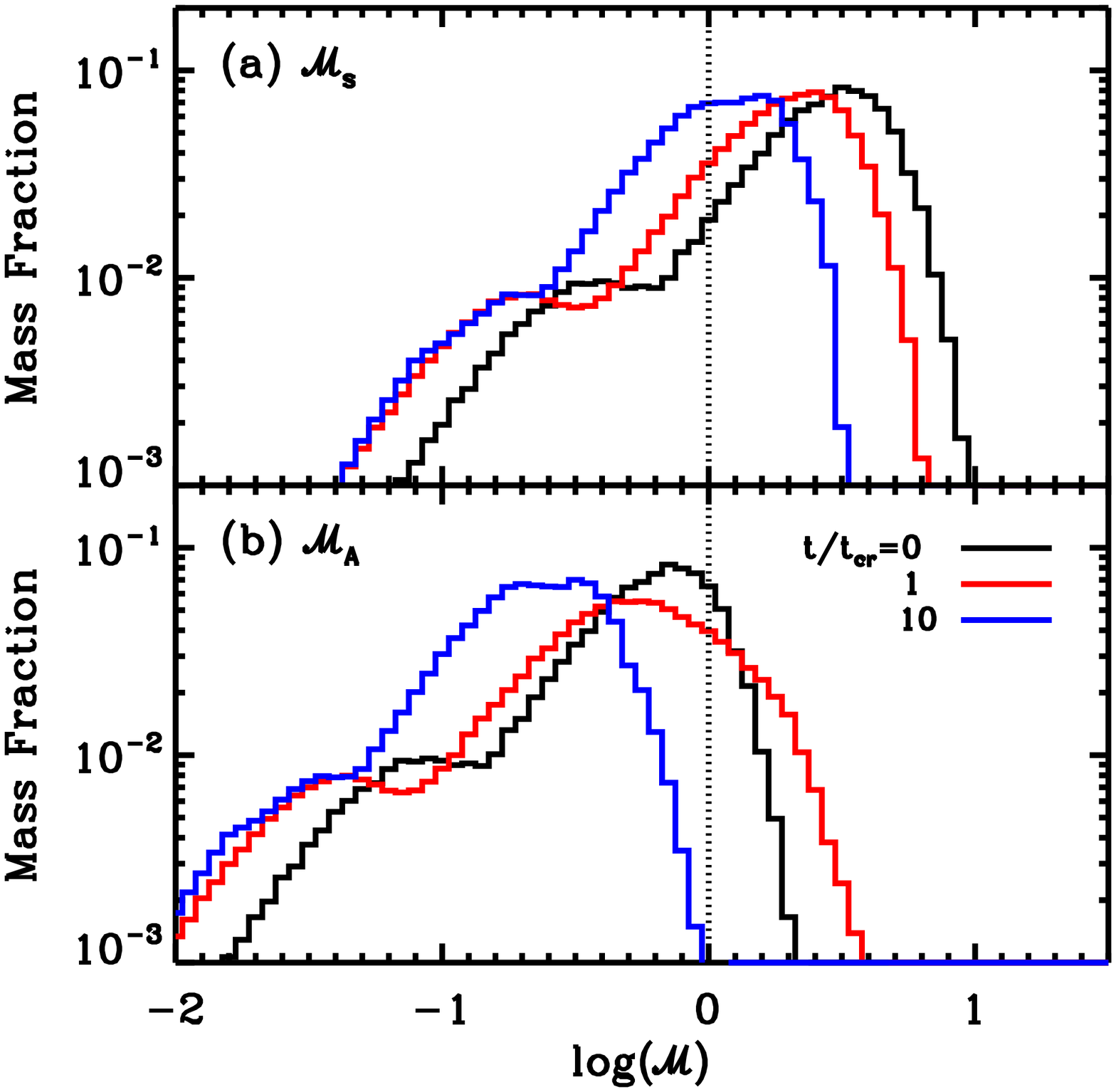}
\caption{(a) Sonic and (b) \Alfven Mach number PDFs for the AS model. The vertical
dotted line denotes a Mach number of unity.  \label{fig:Mpdf}}
\end{figure}

In order to investigate the characteristics of the remaining turbulence
at later times, we begin by analysing the AS model with periodic BCs since it
has a significant amount of energy at later times compared to the others.
Figure~\ref{fig:snap_AS} shows number density snapshots for the AS model in the
XY plane at $z=0$ (\emph{left}), the $\xhat$-$\zhat$ plane at $y=0$ (XZ plane;
\emph{middle}), and the $\yhat$-$\zhat$ plane at $x=0$ (YZ plane; \emph{right})
at $t/\tcross=1$ (\emph{top}) and $10$ (\emph{bottom}).  In the XY plane, it is
clearly shown that there are strong shocks within the CNM sheet at an early
stage due to compressible motions with supersonic velocity (\emph{top-left}),
while the remaining motions are less compressive but vortical
(\emph{bottom-left}). Note that the snapshots at $t/\tcross=1$ may not reflect
the saturated state of turbulence but illustrate a particular choice of our
initial conditions, since the saturated state of turbulence would be achieved
(2-3)$\tcross$ after driving begins \citep[e.g.,][]{sto98}.  Vertical motions are
also induced due to a velocity difference at the contact discontinuity of the
two phases, but they are negligible. The thickness of the CNM sheet is slightly
increased (\emph{middle} and \emph{right}).

For a more quantitative analysis, we decompose the velocity fields into shearing
(divergence-free) and compressible (curl-free) components by using a Helmholtz
decomposition (e.g., \citealt{kow10}).  As seen in the \emph{middle} and
\emph{right} panels of Figure~\ref{fig:snap_AS}, the background density of the
AS model is varying as a function of $z$, and induced vertical motions are
negligible.  A cylindrically-binned PS would thus give good characteristics at
all evolutionary stages for the AS model. This analysis is advantageous to see
how the PS varies along the vertical direction, while it is not suitable to
address anisotropy of turbulence since it shows the $k_z=0$ mode only.  We
first take a 2D FFT in the XY plane for $v_x$ and $v_y$ at a given vertical
position.  The shearing and compressible components of the velocity PS can then
be obtained as functions of $z$ and $k_\perp$ by 
\begin{equation}\label{eq:P2d}
\PS_{\rm sh}(k_\perp;z)=|\kphat\times\tilde{\vel}(\mathbf{k}_\perp;z)|^2 
\quad\textrm{and}\quad 
\PS_{\rm comp}(k_\perp;z)=|\kphat\cdot\tilde{\vel}(\mathbf{k}_\perp;z)|^2,
\end{equation}
where $\tilde{\vel}$ is the velocity field in the Fourier space,
$\mathbf{k_\perp}=(k_x,k_y)$ is the wave vector in the perpendicular (XY)
plane, and $k_\perp=(k_x^2+k_y^2)^{1/2}$ is the cylindrically-binned
wavenumber.  Since our main interests focus on turbulence within the CNM, we
calculate a mass-weighted, vertically-averaged PS for each component as
\begin{equation}\label{eq:Pk}
\abrackets{\PS(k_\perp)}=\frac{\int \overline{\rho}(z)\PS(k_\perp;z)dz}
{\int\overline{\rho}(z)dz},
\end{equation}
where $\overline{\rho}(z)=\int \rho dx dy/(L_xL_y)$ is the
horizontally-averaged density.  We also calculate a specific energy for each
component as a function of $z$:
\begin{equation}\label{eq:Ecomp}
\mathcal{E}(z)=\frac{1}{2}\int \PS(k_\perp;z)\pi k_\perp dk_\perp.
\end{equation}

Figure~\ref{fig:power} plots (a) $\abrackets{\PS_{\rm sh}(k_\perp)}\times
k_\perp^{11/3}$ (\emph{solid}) and $\abrackets{\PS_{\rm comp}(k_\perp)}$
(\emph{dashed}) as a function of $k_\perp$ and (b) $\Ecomp(z)/\Esh(z)$ as a
function of $z$ at $t/\tcross=0$ (\emph{black}), $1$ (\emph{red}), and $10$
(\emph{blue}) for the AS model. In order to show the slope and inertial range
of the PS clearly, we multiply $k_\perp^{11/3}$ in panel (a) to the PS.
Initially, the same amount of energy is assigned in the shearing and
compressible components (see \emph{black} line in
Figure~\ref{fig:power}(b)).\footnote{Previous simulations of decaying
turbulence usually assume purely incompressible initial velocity perturbations
\citep[e.g.,][]{ost99, ost01}.  In our test runs with purely incompressible
initial velocity perturbations, the compressible component is quickly generated
(see also \citealt{ves03}), and the PS of each component is converged in less
than a crossing time to that in models with an initial compressible component.
Although the decay of turbulence is slightly delayed when there is no initial
compressible component, we confirm that the overall evolution remains similar.}
The compressible component keeps decaying rapidly due to strong shock
dissipation, while the shearing component survives longer as seen in
Figure~\ref{fig:power}(a) (see also \emph{left} panels of
Figure~\ref{fig:snap_AS}). As time goes by, the power diffuses to larger scales
via the inverse cascade, and the shearing mode at larger scales starts growing
and finally dominates, while at small scales both the compressible and shearing
components keep decaying. The shearing component of the PS at $t/\tcross=10$
(the final time of simulation) follows a power-law shape with an exponent of
$-11/3$ in an inertial range of $4\le k_\perp L/2\pi\le20$.  The slope is
consistent with the expected value in incompressible hydrodynamic and MHD
turbulence \citep{kol41,gol95}, which is also known to be applicable to \Alfven
modes (as well as slow modes) in compressible MHD turbulence \citep{cho03}.
This slope is achieved and preserved after one crossing time, while the
inertial range varies.  The remaining turbulent energy within the CNM sheet at
later times in the AS model is thus mostly in the form of incompressible
motions and cascades from larger to smaller scales as in traditional
incompressible turbulence. It is not everlasting as in purely thin-disk models
with a potential field outside \citep{bas10}.

It is evident that incompressible motions in the direction perpendicular to the
CNM sheet survive longer and are responsible for the reduced decay rate.
Initial velocity perturbations deform magnetic fields, and a part of the
perturbed kinetic energy is stored in the form of magnetic energy. Since the
\Alfven waves excited by motions in the direction perpendicular to the magnetic
field cannot escape the simulation domain with periodic BCs, they reenter the
CNM sheet and help to preserve shearing motions in the CNM sheet.  After one
\Alfven wave crossing time, the evolution of turbulent energy within the CNM
sheet is affected by the reentering waves and shows a reduced decay rate
(see Figure~\ref{fig:BCs}).

The mean energy ratios between the compressible and shearing components, $\int
\Ecomp(z)dz/\int\Esh(z)dz$, are $16\%$ and $0.8\%$ at $t/\tcross=1$
and $10$, respectively.  However, within the CNM sheet (at around $z=0$),
the energy in the compressible mode remains up to $35\%$ and $6\%$ of
the shearing mode energy at $t/\tcross=1$ and $10$, respectively. At the
final snapshot, the energy ratio at the midplane is about an order of magnitude
greater than at higher-$|z|$. Since the amplitude of incompressible motions is
still supersonic within the CNM sheet (see Figure~\ref{fig:Mpdf}), the more
compressible mode can be more easily generated than in the WNM.  However, the
generation of compressible modes by incompressible \Alfven waves is limited
\citep{cho02,cho03} so that the energy ratio is still very small (only a few
percents) even within the CNM sheet.

Since our model is not isothermal, the probability distribution functions (PDFs) of
sonic and \Alfven Mach numbers provide better understanding of the velocity
dispersions of each gas phase.  The Mach numbers for each gas parcel are defined by
$\mathcal{M}_s\equiv v_{\rm 1D}/c_s$ and $\mathcal{M}_A\equiv v_{\rm 1D}/v_A$,
where $v_{\rm 1D}\equiv [(v_x^2+v_y^2)/2]^{1/2}$, $c_s\equiv(P/\rho)^{1/2}$,
and $v_A\equiv B/(4\pi\rho)^{1/2}$.  Note that we ignore the velocity in the
vertical direction $v_z$, which is negligibly small compared to $v_x$ and $v_y$
at all times in the AS model.  Figure~\ref{fig:Mpdf} plots mass fractions of gas in
the AS model for (a) sonic and (b) \Alfven Mach numbers to put more emphasis on
the CNM.  For $\beta=0.1$, initial mean Mach numbers for the CNM are
$\mathcal{M}_s =\sigma_{\rm 1D,0}/c_c\sim3.3$ and $\mathcal{M}_A=\sigma_{\rm
1D,0}/v_{A,c}\sim0.75$.  However, as seen in Figure~\ref{fig:Mpdf}(b), about
$32\%$ of the CNM by mass is in the super-\Alfvenic regime initially.  Overall
the PDFs gradually move toward smaller values as $v_{\rm 1D}$ decreases.  In
the high density tail of the $\mathcal{M}_A$-PDF, $\mathcal{M}_A$ is initially
increased even though $v_{\rm 1D}$ keeps decreasing. This is due to an initial
lowering of $v_A$ for the densest gas since there are not only horizontal
motions but also vertical compression (see \emph{top} panels of
Figure~\ref{fig:snap_AS}), which enhances density but not magnetic field
energy. At $t/\tcross=10$, $66\%$ of the CNM by mass is still supersonic, but
the entire gas is sub-Alfv\'enic.

\subsubsection{BS and CS Models\label{sec:BSCS}}

\begin{figure*}
\plotone{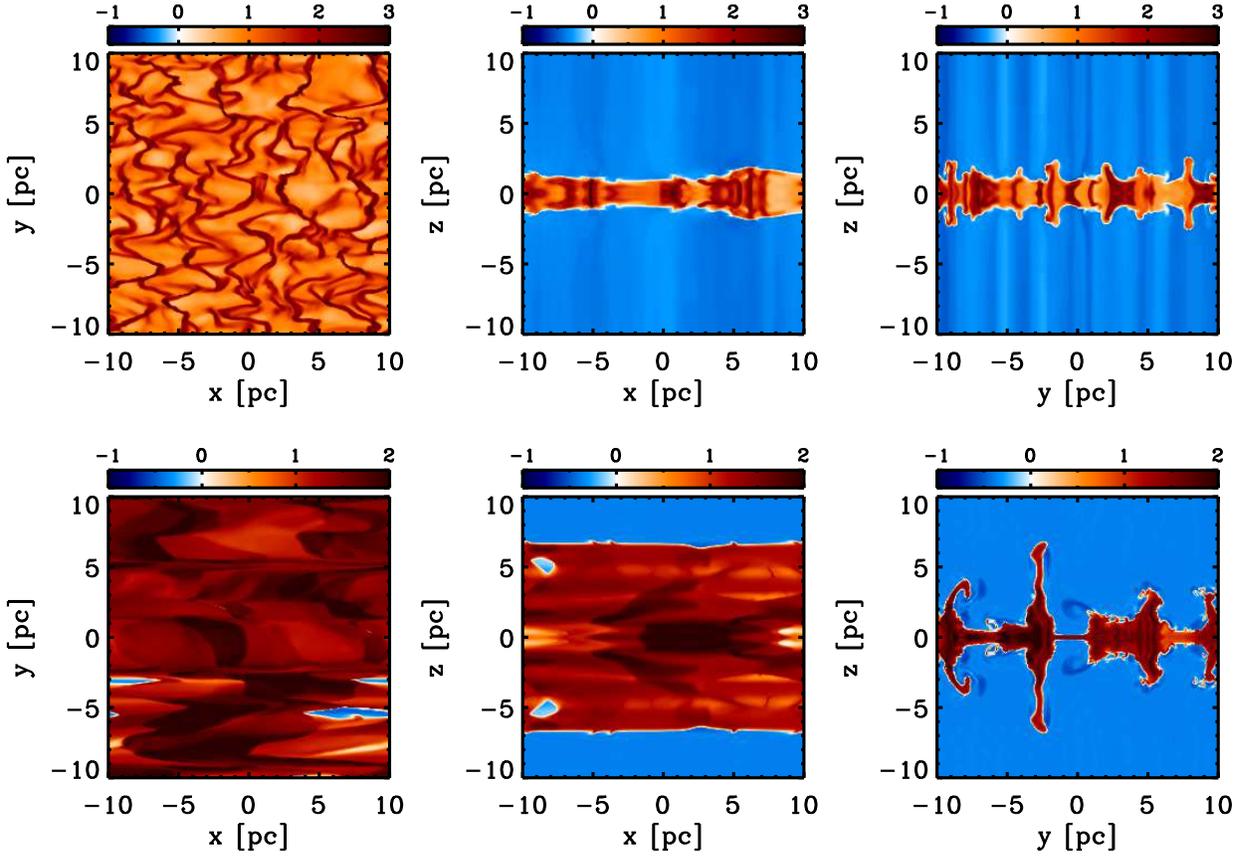}
\caption{Same as Figure~\ref{fig:snap_AS}, but for the BS model. Note that
\emph{middle} panels are slices along $y=-2.9\pc$ (not $y=0\pc$) to show the
structure of the sheet in the XZ plane.
\label{fig:snap_BS}}
\end{figure*}

\begin{figure*}
\plotone{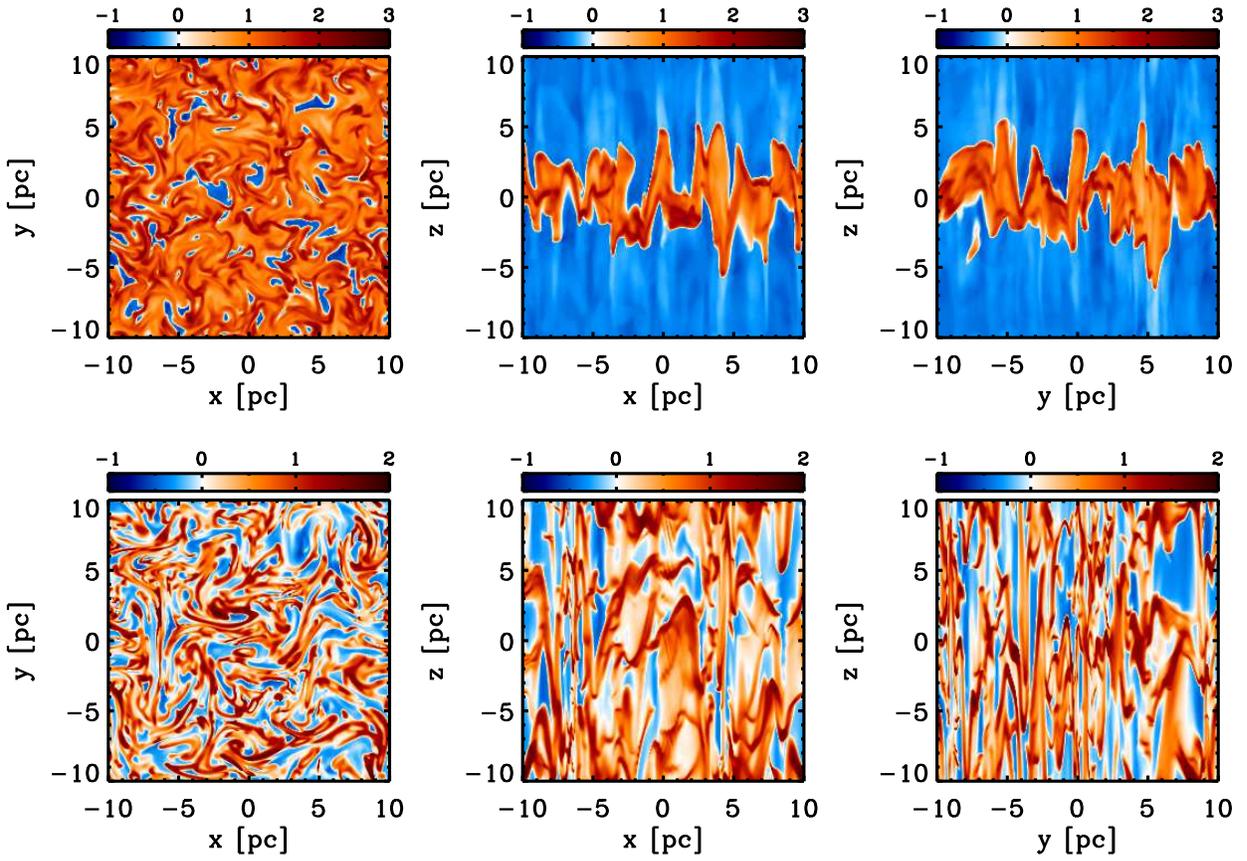}
\caption{Same as Figure~\ref{fig:snap_AS}, but for the CS model.
\label{fig:snap_CS}}
\end{figure*}

\begin{figure}
\plotone{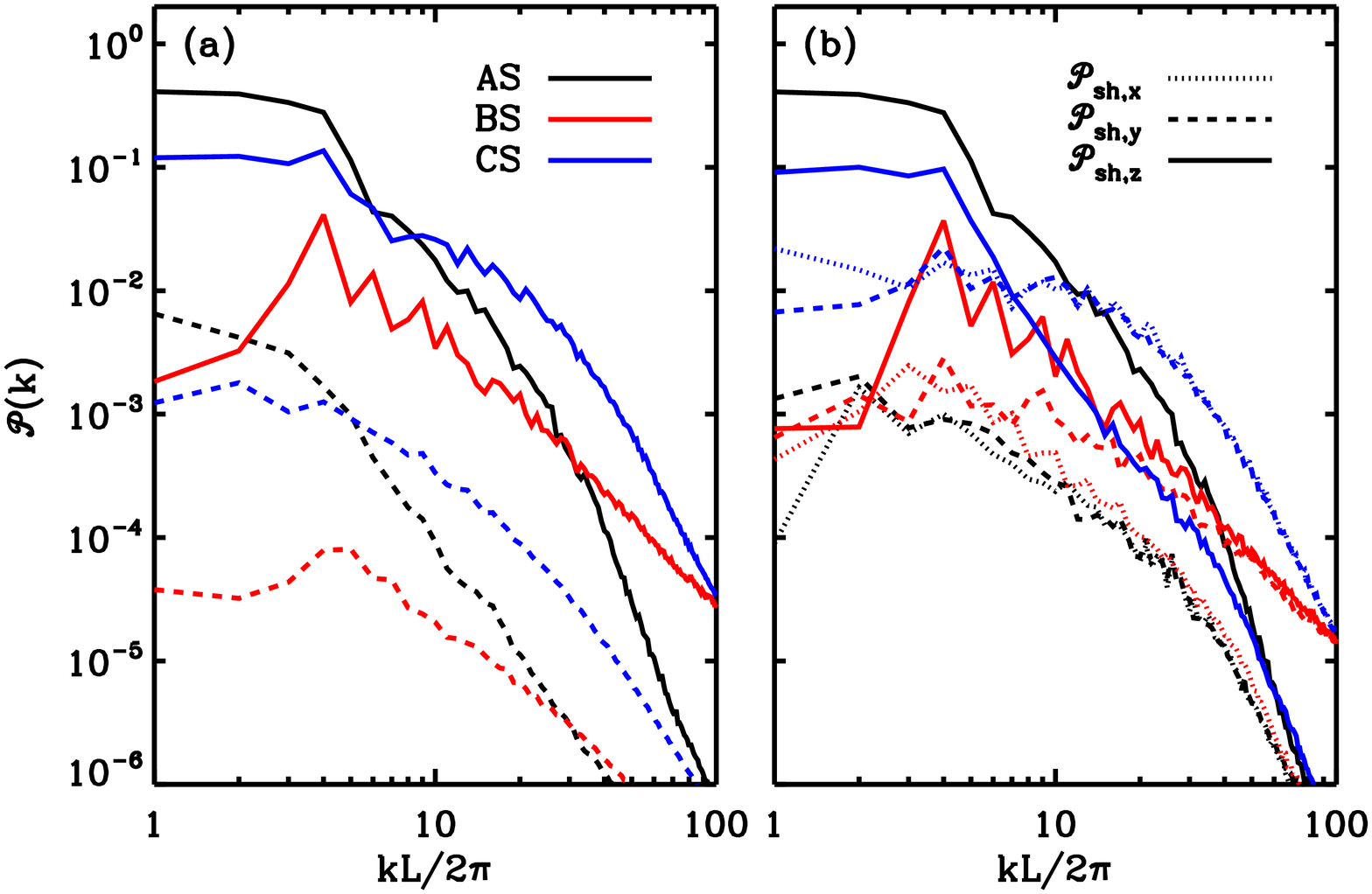}
\caption{(a) Spherically-binned PS for the shearing (\emph{solid}) and
compressible (\emph{dashed}) components for $\beta=0.1$ models at
$t/\tcross=10$. (b) Directionally-decomposed PS for the shearing components.
 \label{fig:PS_all}}
\end{figure}

Figures~\ref{fig:snap_BS} and \ref{fig:snap_CS} show number density slices in
logarithmic color scales for the BS and CS models, respectively, as in
Figure~\ref{fig:snap_AS}.  In contrast to the AS model, the initial CNM sheet is
not maintained, due to non-negligible vertical motions in both models. The
structures in the BS and CS models are greatly different from one another.

Since the gas can flow more easily along the magnetic field lines, the
initial motions in the XY plane for the BS model soon become anisotropic. The
kinetic energies in the $\yhat$- and $\zhat$-directions become comparable after
$t=\tcross$, while the remaining energy, which keeps decaying rapidly (in
contrast to the AS and CS models), is deposited to gas flows in the
$\xhat$-direction.  At $t=\tcross$, filaments in the CNM sheet align to the
$\yhat$-direction in the XY plane, perpendicular to the initial magnetic field
(see \emph{top-left} panel of Figure~\ref{fig:snap_BS}).  Initially formed
filaments keep moving, colliding, and merging along the $\xhat$-direction.
These overdense regions are soon smeared out along the $\xhat$-direction but
preserved in the $\yhat$-direction.  Thus, overdense regions then expand
vertically and form vertical sheets in the XZ plane (see \emph{middle} and
\emph{right} panels of Figure~\ref{fig:snap_BS}).\footnote{The number, shape,
and structure of the sheets may depend on characteristics of the initial
perturbation since vertical motions are mainly induced by the velocity
difference of horizontal motions at the contact discontinuity. For the BH model
(or AH model), there is no specific structure aligned to the $\xhat$-direction,
while the CNM sheet still expands vertically since kinetic energy becomes
comparable in all directions. Note that the detailed structural evolution is
out of the scope of this paper and does not affect the energy decaying rates,
which are the main focus of this paper.}

As we noted in the kinetic energy evolution of Series C (Figure~\ref{fig:tevol_C}),
initial vertical turbulent motions induce an overall expansion of the CNM sheet
due to (turbulent) pressure imbalance. It is clearly seen in the \emph{top} panels
of Figure~\ref{fig:snap_CS}. After about $t/\tcross=2$, the CNM and WNM are mixed up
completely and coexist at all vertical positions. The turbulent energy in the XY
plane decays a bit longer than in the AS model, and incompressible motions
at large scales eventually dominate. In the \emph{bottom} panels of
Figure~\ref{fig:snap_CS}, predominant vortical motions are evident in the XY
plane as in Figure~\ref{fig:snap_AS}, while the vertically stretched CNM is
shown in the XZ and YZ planes.

By comparing the AS and BS models, we see that the initial magnetic field
direction results in drastic differences for turbulent energy and structural
evolution. On the other hand, there are many similarities between the AS and CS
models despite the substantial structural difference in the vertical slices.
In order to characterize turbulence quantitatively, we again investigate the
velocity PS using the Helmholtz decomposition.  For the BS and CS models, the
vertical motions at later times are not negligible, so that the plane-parallel
PS analysis done in Section~\ref{sec:AS} is not suitable. It is better to use
velocity fields in full three dimensions. We decompose the velocity fields in
Fourier space in a similar manner but for a three-dimensional wave vector
$\mathbf{k}=(k_x,k_y,k_z)$ and spherically-binned wavenumber
$k=(k_x^2+k_y^2+k_z^2)^{1/2}$ (see Eq.(\ref{eq:P2d})). There is no additional
averaging needed for a spherically-binned PS. Note that the shearing component
of the PS in three dimensions has three components, representing the power of
incompressible motions perpendicular to the each axis, referred to as $\PS_{\rm
sh,x}$, $\PS_{\rm sh,y}$, and $\PS_{\rm sh,z}$.
 
Figure~\ref{fig:PS_all} plots the spherically-binned PS of the AS
(\emph{black}), BS (\emph{red}), and CS (\emph{blue}) models at $t/\tcross=10$.
In panel (a), the shearing (\emph{solid}) and compressible (\emph{dashed})
components are shown as in Figure~\ref{fig:power}(a). Note that for the AS
model, $\PS_{\rm sh}$ is almost identical to that in Figure~\ref{fig:power}(a),
while $\PS_{\rm comp}$ shows slightly smaller power at large scales (small
$k_\perp$ and $k$). This is because the analysis does not give a mass-weighted
velocity PS, but rather a volume-weighted PS that originates not from the CNM
but from the WNM.  There is a clear signature of the enhanced incompressible
component at large scales for the CS model, while there is no such a feature in
the BS model. The ratios of the compressible to incompressible components of
the integrated specific energy, $\Ecomp/\Esh=\int \PS_{\rm comp}(k)4\pi k^2
dk/\int \PS_{\rm sh}(k)4\pi k^2dk$, are about $1\%$ for all models, again
implying that the compressible component is hardly generated by incompressible
\Alfven waves.

In Figure~\ref{fig:PS_all}(b), we plot the shearing component of the PS
perpendicular to each wave vector. For the AS model, it is clearly seen that
the shearing component perpendicular to the magnetic field, $\PS_{\rm sh,z}$,
is dominant at all scales, justifying our plane-parallel analysis in
Section~\ref{sec:AS}.  For the CS model, $\PS_{\rm sh,z}$ (\emph{blue solid}) is
remarkably similar to that in the AS model with slightly lower power despite a
non-negligible $\PS_{\rm sh,x}$ and $\PS_{\rm sh,y}$ in the CS model.
This again implies a Kolmogorov-like scaling $\PS\propto k_\perp^{-11/3}$
as in \citet{cho03}. Most of energy is  deposited into larger scales
so that slowly decaying kinetic energy at later times is mainly due to the
large-scale shearing component in the direction perpendicular to the magnetic
field, as in the AS model.

The characteristics of turbulence in the BS model are quite different from those
of the AS and CS models, as expected. The shearing component still dominates the
compressible component, but there is no clear enhancement of large scale modes.
For the shearing component, the perpendicular component ($\PS_{\rm sh,x}$) has
less energy than the parallel components, in stark contrast to the AS and CS
models, whose energy is mainly dominated by the perpendicular component
($\PS_{\rm sh,z}$).

\section{Discussion\label{sec:discussion}}

Based on numerous local isothermal simulations of decaying MHD turbulence
during the last two decades, it has been widely accepted that compressible MHD
turbulence in the ISM is decaying rapidly regardless of magnetic field
strength. Due to inherent limitations of local modeling, however, the effects
of global structures of magnetic fields and the diffuse multiphase ISM have not
been explored. Idealized two-dimensional models of thin sheets with
magnetic fields anchored in the interstellar space have shown self-sustained
turbulence driven by the global MHD effect (magnetic-tension driven mode) in a
flux-freezing limit \citep{bas10}.  In this paper, we extend their models to
more realistic interstellar conditions, consisting of thin CNM sheets embedded
in low density WNM \citep{hei03,hei05b}. By solving interstellar cooling and
heating explicitly with a set of ideal MHD equations, we are able to
investigate long-term evolution of turbulent energy decay and characteristics
of the remaining energy in the diffuse multiphase ISM. 

In contrast to the idealized models of \citet{bas10}, who assumed instantaneous
field redistribution toward a potential field outside thin sheets, the kinetic
energy in our models cannot persist (see also \citealt{kud11}). The initial
evolution of the kinetic energy is rather similar to local models characterized by
the decay times shorter than a flow crossing time and power-law shape decay of
$E\propto t^{-1}$ (e.g., \citealt{mac98,sto98}).  For our strong field model
(AS model) with vertical magnetic fields and anisotropic turbulence (no initial
vertical motions), we find a clear signature of reduced decay rate at later
times, and the kinetic energy remains at $32\%$ and $18\%$ of the
initial energy at $t/\tcross=1$ and $10$, respectively. Similar results
are found in the weaker field models, but less energy remains for weaker
magnetic fields (Figure~\ref{fig:tevol}(a)). This dependence on the field
strengths is also shown in Series C (with vertical fields and isotropic
turbulence), while turbulent energy keeps decaying as a power-law of $E\propto
t^{-1}$ in Series B (with horizontal fields and anisotropic turbulence)
irrespective of magnetic field strengths (Figure~\ref{fig:tevol}(b) and (c)).
In \citet{ost99}, two-dimensional local models with horizontal magnetic fields
have also reported reduced decay rates for strong magnetic fields, arising
mainly from a lack of dissipation in the symmetric direction (see also
\citealt{gam96,kud03}).  Although the initial velocity fields are
two-dimensional in Series A and B, vertical motions emerge naturally since
there is no artificial suppression of dissipation in the symmetric direction.
The reduced decay rates shown in the AS and CS models are indeed due to the
survival of incompressible motions in the direction perpendicular to the
magnetic field (see Section~\ref{sec:turb_chr}). 

We use the Helmholtz decomposition method to obtain incompressible and
compressible components of the velocity PS, providing detailed characteristics
of the remaining turbulence. At later times, almost all the energy in Series A
and C is in the form of shearing motions perpendicular to the magnetic field
(in the XY plane) at the largest scale (Figure~\ref{fig:PS_all}). The remaining
energy in the compressible component has only $1\%$ of the incompressible
component at the end of simulation ($t/\tcross=10$), while the CNM sheets
contain more compressible energy than the WNM (Figure~\ref{fig:power}).
Although a significant fraction of the CNM by mass is in supersonic regimes,
incompressible \Alfven waves are hardly generating compressible motions,
implying a weak coupling of MHD waves in the absence of driving (c.f.,
\citealt{cho02,cho03}). The velocity PS for the AS and CS models at later times
show power-law shapes $\PS_{\rm sh} \propto k^{-11/3}$, close to that in
classical incompressible turbulence \citep{kol41,gol95}.  This is not
surprising since \Alfven modes (and slow modes as well) follow a
Kolmogorov-like spectrum \citep{cho02,cho03}, and these are the dominant modes
in our simulations.

Based on our models, we find for the AS model (and also the CS model) a
significant delay of turbulent decay and non-negligible remaining turbulent
energy ($v_{\rm 1D}\sim1\kms$), which may be associated with supersonic
non-thermal linewidths within the CNM. This requires a sheet-like CNM
distribution with strong magnetic fields threading perpendicular to the sheet,
as well as periodic BCs in the vertical direction.
For a typical thermal pressure of the diffuse ISM,
$P_0/\kbol\sim3-4\times10^3\Punit$ in the Solar neighborhood \citep{jen01,
jen11, wol03}, a strong magnetic field assumed in the AS model ($\beta=0.1$)
implies $B_0\sim10-12\muG$ (see Eq.(\ref{eq:B0})).  The observed average
strength of the total magnetic field in spiral galaxies is about $10\muG$,
while weaker fields ($\sim 5\muG$) are observed for radio-faint galaxies in the
Local group, and star-forming galaxies show stronger field strengths of
$20$-$30\muG$ (see \citealt{bec05} and references therein).  The median value
of the CNM ($10\pcc<n_{\rm H}<300\pcc$) in the Milky Way is
about $6\muG$ \citep{hei05b, cru10, cru12}.  Thus the typical magnetic field
strengths in the ISM are generally expected to be in the strong-field regime
with respect to the thermal pressure, as in the AS model.

A sheet-like CNM perpendicular to the local magnetic field is also expected
to be common. One direct formation mechanism for CNM structure is condensation
of the WNM due to supersonic colliding flows driven by e.g., supernova
explosions.  Recent simulations of supersonic colliding flows with magnetic
fields by \citet{hei09} show that a thin CNM sheet perpendicular to the field
can be formed when the WNM flows along a strong magnetic field, otherwise
sheets are soon disrupted by the nonlinear thin shell instability
\citep{vis94,hei07}.  In numerical simulations of structure formation in the
multiphase ISM solely by thermal instability, CNM filaments or sheets are
usually formed along the direction perpendicular to the field lines since gas
flows across the field are prevented, especially in strong field cases (e.g.,
\citealt{hen00,ino12}).  When CNM clouds are formed by large-scale gravitational
collapse within slightly supercritical gas, planar structures perpendicular to
the magnetic field direction are also expected to form first. Subsequently,
self-gravity overwhelms other forces, such as magnetic-tension, in order to
cause isotropic collapse. In a high-resolution \ion{H}{1} absorption line
survey using the Arecibo radio telescope, \citet{hei03} have suggested ``blobby
sheets'' to characterize the structure of CNM clouds rather than isotropic
clouds (see also \citealt{gib00,mey06}), while the orientation of the magnetic
fields is still uncertain \citep{hei05b}.

Such commonly observed CNM sheets imply that the CNM and WNM should be in total
pressure equilibrium in direction perpendicular to the sheet, otherwise the
sheet is dispersed easily as seen in Series C (see Figure~\ref{fig:snap_CS}).
The total pressure equilibrium can be expressed as
\begin{equation}\label{eq:pequil}
P_{\rm th,w}+P_{\rm turb,w}+\delta P_{\rm mag,w}=
P_{\rm th,c}+P_{\rm turb,c}+\delta P_{\rm mag,c},
\end{equation}
where $P_{\rm th}=\rho c_s^2$, $P_{\rm turb}=\rho \sigma_z^2$, and $\delta P_{\rm
mag}=\delta B^2/8\pi$  are the thermal, turbulent, and turbulent magnetic
pressures, respectively. Subscripts `w' and `c' represent the WNM and CNM, respectively.
Assuming that the density, temperature, vertical velocity dispersion, turbulent
magnetic fields are constant within each gas phase, Eq.(\ref{eq:pequil})
results in a thermal pressure ratio  between the two phases 
\begin{equation}\label{eq:pratio}
\mathcal{R}_p\equiv\frac{P_{\rm th,w}}{P_{\rm
th,c}}=\frac{1+\mathcal{M}_c^2+\beta_c^{-1}}{1+\mathcal{M}_w^2+\beta_w^{-1}}.
\end{equation}
The left hand side of Eq.(\ref{eq:pratio}) can be constrained by
thermal equilibrium.  For given interstellar cooling and heating rates, there are
maximum and minimum thermal pressures of the WNM and CNM, $P_{\rm max}$ and $P_{\rm
min}$, respectively, and $P_{\rm max} \sim 3 P_{\rm min}$ \citep{wol95,wol03}.
Since the cooling time scale is generally short compared to the gas
dynamical time scales, gas is expected to be in thermal equilibrium even in the
case of vigorously turbulent disks (e.g., \citealt{pio05,kko11}).  Thus, it is
likely to have $\mathcal{R}_p<3$. More generally, the CNM can have higher thermal
pressure due to self-gravity (e.g., \citealt{jen11}) so that it is more natural
to have $\mathcal{R}_p\simlt1$.

Non-thermal linewidths in the CNM and WNM are observed in the ranges of $\sim2-7\kms$
and $\simgt10\kms$, respectively, resulting in $\mathcal{M}_c>3$ and
$\mathcal{M}_w\sim1$-$2$ \citep{hei03,kal09}. If the magnetic field is uniform and
perpendicular to the sheet, we can neglect a contribution from magnetic pressure.
We then have the thermal pressure ratio $\mathcal{R}_p>5$ for $\mathcal{M}_c>3$
and $\mathcal{M}_w\sim1$ from the observed velocity dispersions, which is
inconsistent with the expectation from thermal equilibrium of the multiphase
ISM. If the turbulence within CNM sheets is anisotropic so that the turbulent
pressure in the parallel direction to the field is negligible as in Series A and B,
$\mathcal{R}_p$ can be kept around unity with the observed Mach numbers of each
medium. There are also other ways to explain this contradiction. If
CNM sheets are confined mainly by the ram pressure of the WNM as in colliding flow
simulations, an additional term appears in denominator of Eq.(\ref{eq:pratio}),
reducing $\mathcal{R}_p$ if the Mach number of the colliding flow is similar to
$\mathcal{M}_c$ (e.g., \citealt{hei09}). It is also possible to have isotropic
turbulence in sheets if the turbulent magnetic pressure overwhelms other terms and
$\beta_c\simgt\beta_w$. Although the turbulent magnetic field can be as strong
as the uniform field \citep{hei05b,han06}, anisotropic turbulence is still
preferred since $\beta\sim0.3$ is not enough to overwhelm turbulent terms.

As we have shown in Section~\ref{sec:turb_evol_BC}, an isolated CNM sheet
(modelled by continuous BCs) cannot maintain the turbulent energy for a longer
time. Instead, the energy decays rapidly as in local simulations. The reduced
decay rate found in the AS model is thus a consequence of the replenishment of
\Alfven waves into the CNM sheet caused by the periodic BCs. In the real ISM, the
CNM sheets would be neither placed periodically in the direction perpendicular
to the magnetic field nor would they be isolated. Realistic global simulations
that model larger magnetized, multiphase \ion{H}{1} clouds containing a number
of CNM sheets would be necessary to further investigate the reduced decay
rate of turbulence found in the AS model.

In this paper, we focus on the long-term evolution of decaying turbulence under
the influence of interstellar cooling and heating with global structures of
magnetic fields. In reality, however, dynamical processes whose time scales are
shorter than the decay time of turbulence should be taken into account to
explore dynamical and structural evolution of the multiphase ISM. Using a
driving scheme in Fourier space, several studies have explored characteristics
of turbulence in the multiphase ISM, such as density and pressure PDFs, mass
fractions, structure functions, and power spectra (e.g.,
\citealt{gaz05,gaz10,gaz13,sei11,sau13}). More realistic situations have also
been considered with driving sources of turbulence such as converging flows
(e.g., \citealt{aud05,aud10,hen07,hei05,hei09,vaz07}) and propagating shock
waves (e.g., \citealt{koy00,koy02}). Since only a few of them considered the
effect of magnetic fields (e.g, \citealt{gaz09,hei09,hen00,hen06}), however,
many MHD turbulence characteristics within the multiphase ISM remain
unexplored.  Especially, driven turbulence within CNM sheets with self-gravity
can be investigated in our future work.

\section{Summary\label{sec:summary}}

Using three-dimensional ideal MHD simulations with interstellar cooling and
heating, we have shown that the long-term evolution of decaying turbulence in CNM
sheets embedded in a WNM can be different from the usual results of local isothermal
simulations. Both models share common characteristics of decaying turbulence
described by power-law decay with short decay time at early evolutionary
stages.  For strong and perpendicular (to the sheet) magnetic fields
however, turbulent kinetic energy decays slowly at later times and can survive
longer until $10\tcross$ in the form of incompressible motions in the plane of
the sheet. A significant amount of energy, $\sim20\%$ of the initial
energy, remains in models with strong fields ($\beta=0.1$) and anisotropic
velocity perturbations (no initial vertical motions; AS model). One
more very important requirement for a longer-lived turbulence maintained by a
strong magnetic field is a mechanism to compensate for the leaking of \Alfven
waves (periodic BCs in our models).  While idealized, such BCs may represent
some aspects of the interaction of multiple CNM sheets within
the observed neutral ISM \citep{hei05b}.

\acknowledgments{
The authors are grateful to the referee for a constructive report on the
manuscript, and also to Drs. Jongsoo Kim and Eve Ostriker for helpful
discussions on boundary conditions.  This work was made possible by the
facilities of the Shared Hierarchical Academic Research Computing Network
(SHARCNET: www.sharcnet.ca) and Compute/Calcul Canada. CGK is supported in part
by a CITA National Fellowship. SB is supported by a Discovery Grant from NSERC.
}

%\bibliographystyle{apj}
%\bibliography{../../ref}{}

\end{document}